\documentclass[preprint]{aastex}

\newcommand \kms          {km~s$^{-1}$}
\newcommand \ha           {H$\alpha$}
\newcommand \hb           {H$\beta$}
\newcommand \oiii         {[\ion{O}{3}]}
\newcommand \nii          {[\ion{N}{2}]}

\newcommand \sii          {[\ion{S}{2}]}

\begin{document}

\title{Narrow Lines in Type II Supernovae -- Probing the Circumstellar 
Nebulae of the Progenitors}
\shorttitle{Narrow Lines in Type II Supernovae}

\author{Robert A. Gruendl, You-Hua Chu\altaffilmark{1,2}}
\affil{Astronomy Department, University of Illinois,
1002 West Green Street, Urbana, IL 61801}
\email{gruendl@astro.uiuc.edu, chu@astro.uiuc.edu}
\author{Schuyler D. Van Dyk\altaffilmark{1}}
\affil{IPAC/Caltech, Mail Code 100-22, Pasadena, CA 91125}
\email{vandyk@ipac.caltech.edu}
\author{Christopher J. Stockdale\altaffilmark{3}}
\affil{Department of Physics and Astronomy, University of Oklahoma, 
440 West Brooks, Room 131, Norman, OK 73019}
\email{stockdal@rsd.nrl.navy.mil}
\altaffiltext{1}{Visiting astronomer, Kitt Peak National Observatory}
\altaffiltext{2}{Visiting astronomer, Cerro Tololo Inter-American Observatory}
\altaffiltext{3}{Current address: Naval Research Laboratory, NRL-Code 7213,
Washington, DC 20375}


\begin{abstract}

We have carried out a high-dispersion (R$\sim$30,000)
echelle spectroscopic survey of 16 Type II supernovae 
(SNe) to search for narrow emission lines from 
circumstellar nebulae ejected by their massive 
progenitors.  Circumstellar nebulae, if detected, 
provide invaluable opportunities to probe SN 
progenitors.  Of the 16 SNe observed, SN ejecta are
clearly detected in 4 SNe and possibly in another 2 SNe,
interstellar gas is detected in 12 SNe, and circumstellar
material is detected only in SN\,1978K and SN\,1998S. 
In the case of SN\,1978K we are able to place an
upper limit of $\sim$2.2 pc for the size of the 
circumstellar ejecta nebula and note that this is more
consistent with the typical sizes observed for ejecta
nebulae around luminous blue variables rather than 
Wolf-Rayet stars.  In the case of SN\,1998S, our 
observations of the narrow lines $\sim$1 year after the SN 
explosion show variations compared to early epochs.
The nebular lines we observe from SN\,1998S originate from
either the low--density, outer region of a circumstellar
nebula or have become dominated by an interstellar component.

\end{abstract}


\newpage
\section{Introduction}

Supernovae (SNe) of Types Ib, Ic, and II are believed to have massive 
progenitors, because they have been found frequently in or near spiral 
arms and \ion{H}{2} regions but not in elliptical galaxies \citep*{VDHF96}.
Few SNe (e.g., SN\,1987A) have recognized massive progenitors; 
consequently, little is known about the stellar evolution immediately 
before the SN explosion.  

Evolved massive stars are known to undergo copious mass loss,
forming circumstellar nebulae.  
The rings around SN~1987A are an example of such a nebula
\citep{Buetal95}.  The chemical composition and kinematics of
the rings have provided essential constraints that lead to the 
hypothesis that the B3I progenitor Sk\,$-$69$^\circ$202 was a binary 
\citep{Po92}.  These circumstellar nebulae can also be detected
from the presence of narrow H$\alpha$ emission lines 
(FWHM $\leq$ 200 km s$^{-1}$) in spectra of Type IIn SNe 
\citep{Sc90,filip91,filip97}.  The circumstellar nebulae of distant 
SNe cannot be resolved spatially, but their expansion, physical conditions,
and chemical enrichment can be investigated by high-dispersion 
(R $\geq$ \ 30,000) spectra, as recently demonstrated for 
SN\,1997ab \citep{sal98} and SN\,1978K \citep{Chetal99}.

Most available spectroscopic observations of SNe have been made with low
or intermediate spectral dispersion to study the broad spectral lines of
SN ejecta.  These spectra are useful for revealing the presence
of narrow nebular lines, but are not adequate for analysis of nebular 
kinematics and physical conditions.  Therefore, we have undertaken a 
high-dispersion spectroscopic survey of 16 Type II SNe.  The results 
are reported in this paper.

\section{Observations}

High-dispersion spectra were acquired with the echelle spectrograph
using the Kitt Peak National Observatory (KPNO) 4m telescope in March 
1999.  The 79 line mm$^{-1}$ echelle grating was used in combination with
a 226 line mm$^{-1}$ cross-disperser and the long-focus red camera to 
achieve a reciprocal dispersion of 3.5 \AA\ mm$^{-1}$ at \ha .  The 
spectra were imaged with the T2KB CCD detector, where the pixel size 
of 24 $\mu$m corresponds to 0\farcs 24 pixel$^{-1}$ along the slit and  
$\sim$3.7 km~s$^{-1}$ pixel$^{-1}$ at the \ha\ line along the dispersion 
axis.  The spectral coverage was roughly 4000--7000 \AA .
The instrumental resolution for the 1\arcsec\ wide slit,
as measured from the widths of sky lines, was $\sim$10 \kms (FWHM) at \ha.

The sky conditions were moderate to poor due to variable high cirrus and 
nearly full lunar phase, making the target acquisition difficult at times.
The spectra were reduced using the echelle and longslit packages in IRAF
with careful attention paid to the orders that contained \ha\ emission. 
Observations of the spectrophotometric standard GD\,140 were used to
correct for the blaze, illumination, and spectral response, but the 
resulting spectra were not calibrated to an absolute flux scale as the
observing conditions were not photometric.  

In addition to the objects observed at KPNO, we have also observed
SN\,1978K with the echelle spectrograph on the 4m telescope at the
Cerro Tololo Inter-American Observatory (CTIO) in December 2000.  
The instrument was configured with similar optical elements as were used
for the KPNO observations, and the spatial and spectral parameters are
therefore similar.  All observations are summarized in 
Table~\ref{tab_obssum}.

\section{Results}

Our high-dispersion echelle observations are useful in distinguishing 
among three sources of line emission in distant SNe: \begin{itemize}
\itemsep-0.1cm
\item SN ejecta, which can be identified by a broad (width $>$ 1000 
      km~s$^{-1}$), spatially-unresolved H$\alpha$ line; 
\item Unshocked circumstellar material (CSM) ejected by the SN progenitor, 
      which shows
      narrow (width $\le$ 200 km~s$^{-1}$), spatially-unresolved H$\alpha$ 
      and \nii\,$\lambda\lambda$6548, 6583 lines; and 
\item Interstellar medium (ISM) of the host galaxy, which is characterized
      by narrow, spatially-extended H$\alpha$ emission and forbidden
      lines such as \oiii\,$\lambda$5007 and \sii\,$\lambda\lambda$6716,6731.
\end{itemize}
To search for emission from the SN, CSM, and ISM in each of our spectra,
we used the spatio-kinematic descriptions of line emission given above
along with the criterion that the statistical significance of the peak of 
an emission component be $>$3$\sigma$ per resolution element 
(e.g., $\sim$10 \kms $\times$ 1\arcsec\ at H$\alpha$) with respect to the
background rms noise.  The results are summarized in Table~\ref{tab_observe}.

Of the 16 SNe observed, we clearly detect broad line emission from the SN 
ejecta from four sources: SN\,1978K, SN\,1995N, SN\,1997eg, and SN\,1998S.  
In two further cases, SN\,1999E and SN\,1999Z, broad \ha\ emission line 
may be detected, but significant portions of the line profiles are lost
between echelle orders.  It is difficult to distinguish between the 
line and continuum emission components from the SN ejecta of these two
latter objects.

Our low detection rate of the SNe is likely caused by the
faintness of the SNe and the poor sky condition.  The majority of the
SNe observed were not visible in the images taken by the acquisition camera.
Blind offsets from nearby stars had to be employed for target acquisition.  
If the coordinates of a SN from the literature were off by more than 
1\arcsec\ perpendicular to the slit, our blind-offset observations would 
have missed the target entirely.  

\begin{figure}
\figurenum{1a}
\centerline{\plotone{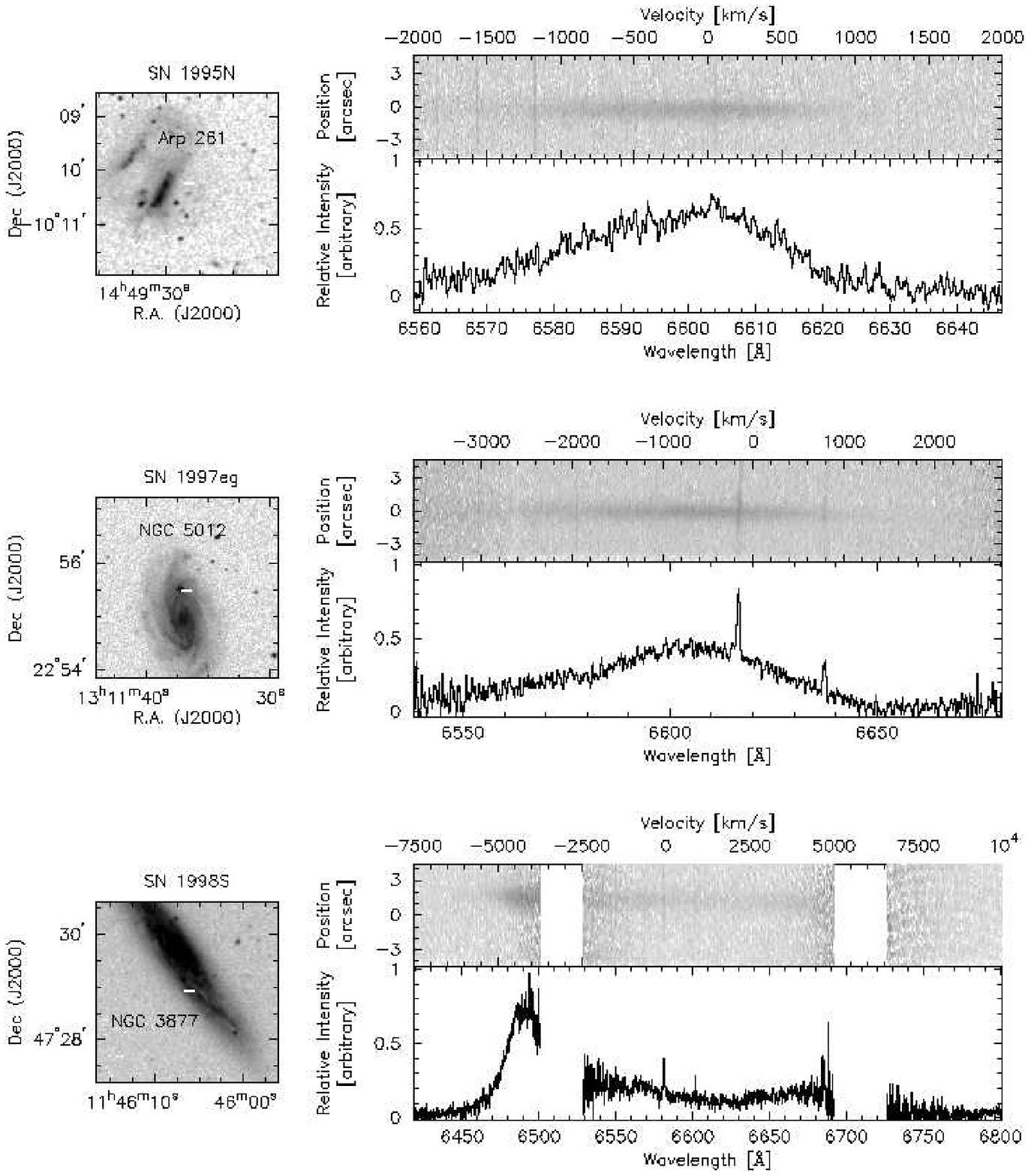}}
\caption{Finding charts and \ha\ spectra of SN\,1995N, SN\,1997eg, and
SN\,1998S. ({\it Left column}) Digitized Sky Survey image of the host 
galaxy/environment for each SN.  The slit position for the echelle 
observations is overlaid as a line segment.  ({\it Right column}) The 
echellogram and the \ha\ line profile for each SN.  The velocity scale
at the top of each panel is relative to the systemic velocity of 
the galaxy.  In some spectra gaps appear; these are cases where multiple
echelle orders contain \ha\ emission from the SN ejecta.}
\label{fig_sndetecta}
\end{figure}

The H$\alpha$ echellograms and line profiles of SN\,1995N, SN\,1997eg, 
SN\,1998S, SN\,1999E, and SN\,1999Z are presented in 
Figure~\ref{fig_sndetecta}\&b, along with finding 
charts.  Spectral scales are given in both observed wavelengths and 
velocities relative to the host galaxy's systemic velocity.  
The broad H$\alpha$ emission from the SN ejecta, detected over
several thousand \kms, is evident in each image.

\begin{figure}
\figurenum{2}
\centerline{\plotone{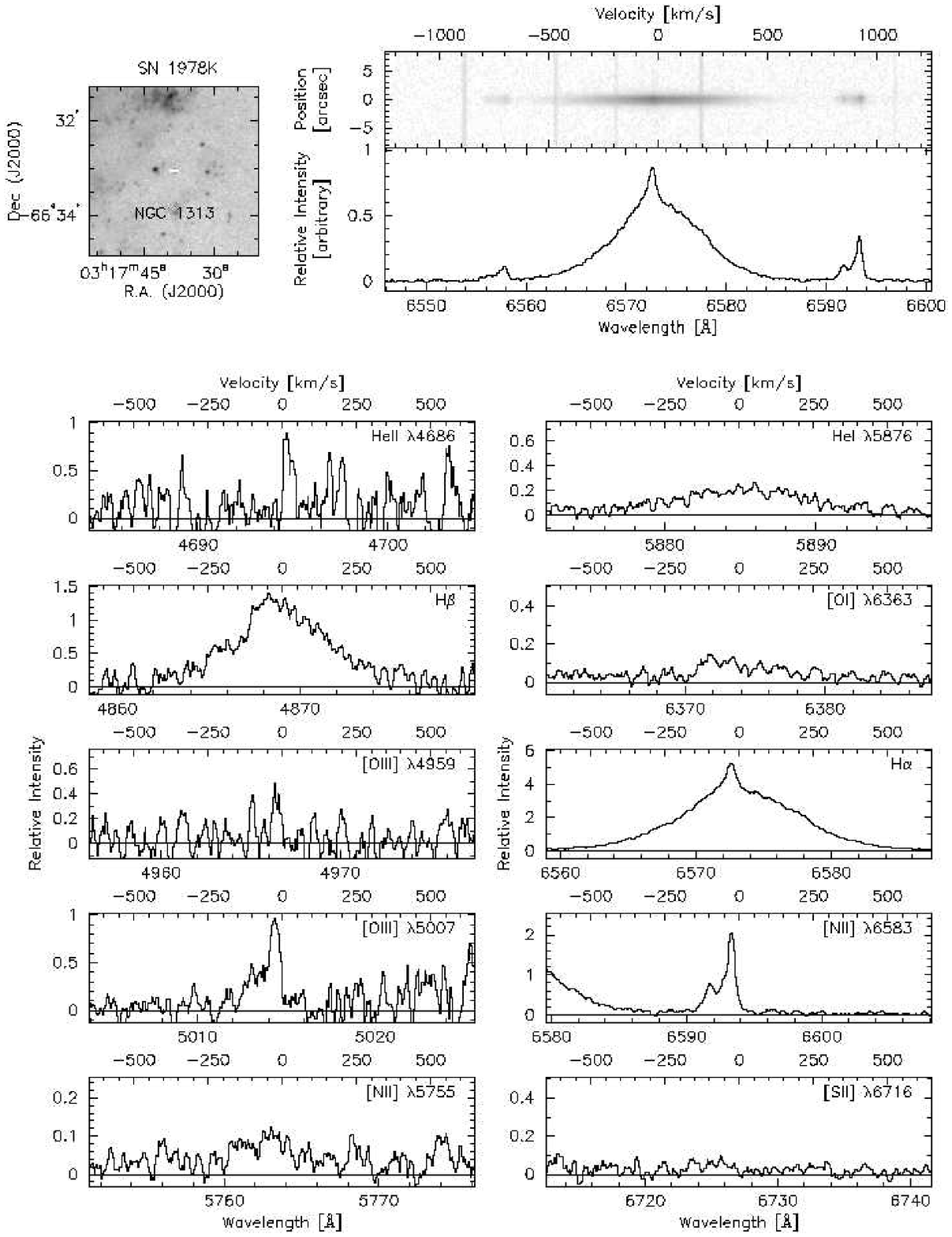}}
\caption{({\it Top}) Same as Figure~\ref{fig_sndetecta} for SN\,1978K.
({\it Bottom}) Profiles of ten spectral lines from 
SN\,1978K that were included in the echelle observations.}
\label{fig_sn1978k}
\end{figure}

SN\,1978K was detected with the highest signal-to-noise (S/N) ratio; several
lines were detected.  Its H$\alpha$ line image and the profiles of 10 spectral 
lines are presented in Figure~\ref{fig_sn1978k}.  While H$\alpha$, 
H$\beta$, and \ion{He}{1}\,$\lambda$5876 lines of the SN ejecta are broad,
they are not as broad as the H$\alpha$ line of the other three SNe we detected.
More interestingly, a narrow nebular component is detected in H$\alpha$,
[\ion{N}{2}] $\lambda\lambda$6548, 6583 and [\ion{O}{3}] $\lambda$5007 
lines.  The high [\ion{N}{2}]/H$\alpha$ ratio indicates that this narrow
component originates from a circumstellar nebula \citep{Chetal99}.

\begin{figure}
\figurenum{3}
\centerline{\plotone{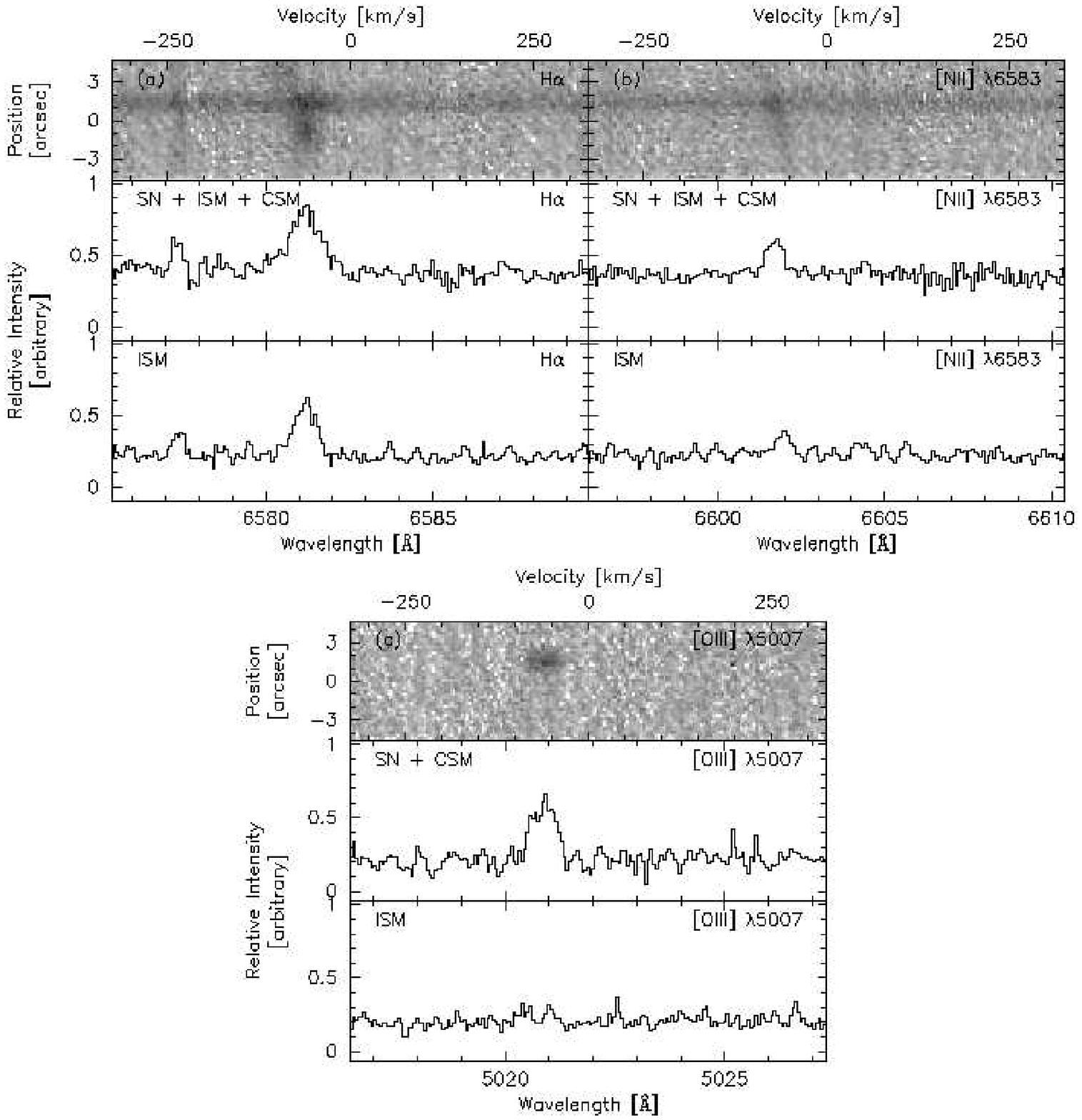}}
\caption{Narrow emission line components in the spectrum of SN\,1998S near
the systemic velocity of the host galaxy.  ({\it Top left panel}) shows 
the \ha\ echellogram and the extracted spectrum on and off the SN.
({\it Top right panel}) shows similar details for the \nii\ $\lambda$6583 line.
({\it Bottom panel} shows similar details for the \oiii\ $\lambda$5007 line.}
\label{fig_sn1998S}
\end{figure}

Besides SN\,1978K, the only other SN in our sample that shows emission
components from a circumstellar nebula is SN\,1998S.  
In Figure~\ref{fig_sn1998S} we show the narrow \ha, \nii, and \oiii\ 
components detected in this spectrum. 
The identification of a circumstellar component in SN\,1998S is not as 
straightforward as that for SN\,1978K because of the presence of a
bright interstellar component.
Details of this identification are given in \S3.5 below.

\begin{figure}
\figurenum{4}
\centerline{\plotone{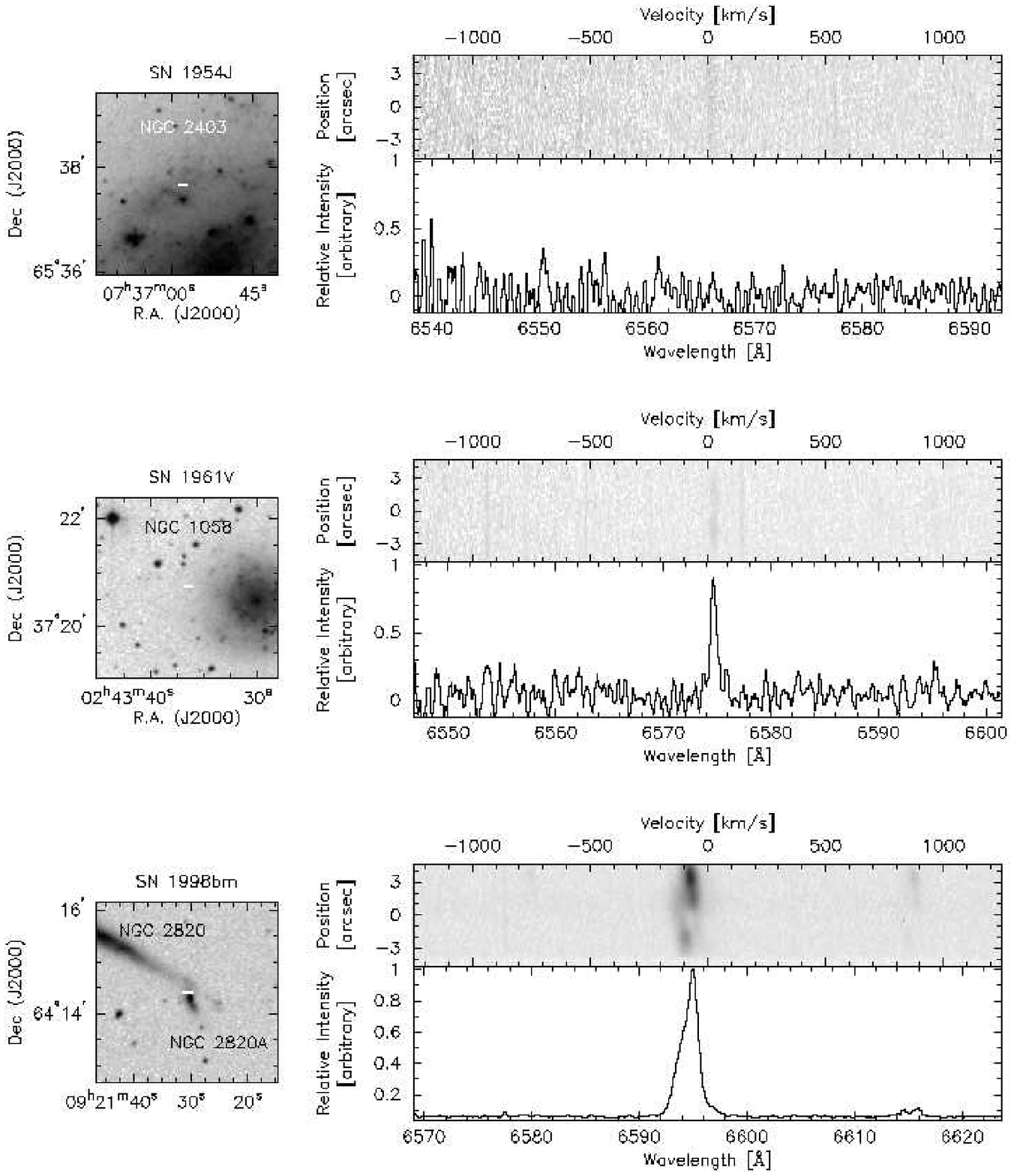}}
\caption{Same as Figure~\ref{fig_sndetecta} for SN\,1954J, SN\,1961V, and
SN\,1998bm.}
\label{fig_nosn}
\end{figure}

Extended interstellar emission is detected in our observations of 12 
SNe.  This high detection rate (12 out of 16) is fully consistent with
their being Type II SNe with massive progenitors.  Three examples of 
interstellar H$\alpha$ emission are presented in Figure~\ref{fig_nosn}.
The interstellar H$\alpha$ line, compared to the telluric OH and H$\alpha$
lines, is broader and shows velocity variations along the slit.

In the remainder of this section we summarize the observed properties 
of individual objects that are interesting.

\subsection{SN\,1954J and SN\,1961V}

SN\,1954J and SN\,1961V, in NGC\,2403 and NGC\,1058 respectively, had spectral
properties similar to a Type II SN,
but were peculiar in four respects: 1) the linewidths of SN\,1961V correspond
to an expansion velocity of only 2000 \kms, 2) the progenitors were 
known, 3) they were underluminous, and 4) their SN light curves were 
followed in the optical for up to eight years.  It has been suggested that 
SN\,1954J and SN\,1961V are luminous blue variables (LBVs) similar to $\eta$~Car
\citep{humphreys94,goodrich89}.  

In optical and near-infrared imaging observations more recent than our
echelle observations, \citet*{smith01} have identified a faint red star with
a position consistent with V12, the progenitor of SN\,1954J.  Similarly, using 
{\it Hubble Space Telescope} ({\it HST}) WFPC1 images of SN\,1961V, 
\citet{filip95} have tentatively identified a faint red star with a position 
consistent with SN\,1961V.  In either case, these red sources could be a 
post-eruption LBV, but neither observation can rule out that these sources 
are an unrelated red supergiant.  Spectroscopic observations similar to 
those reported in this paper are needed.  
Our observations of SN\,1954J and SN\,1961V were based on the best coordinates
for the SNe that were available.  While no emission from a SN or a post eruption 
LBV are detected in these spectra, interstellar emission red-shifted by 
$\sim$ 20 \kms\ with respect to the systemic velocity of each host galaxy 
are present in the spectra.  The coordinates of the candidate 
post-eruption LBV tentatively associated with SN\,1954J \citep{smith01} 
place it more than 2\arcsec\ from our 1\arcsec\ echelle slit center and 
therefore our observations do not include significant emission from this 
source.

Optical \ha$+$\nii, \oiii, and \sii\ images of the vicinity of SN\,1961V
show two \ion{H}{2} regions 
separated by $\sim$3\arcsec\ \citep{fesen85} that are coincident with two
non-thermal radio sources \citep*{branch85,cowan88}.  The eastern source
is roughly coincident with SN\,1961V.  Recent VLA observations at 6~cm
and 18~cm by \citet{stockdale01} confirm that both radio sources are 
non-thermal and show that SN\,1961V's radio emission has decayed.  These
observed properties are consistent with many known SNe that have been
recovered at radio wavelengths.

\begin{figure}
\figurenum{5}
\centerline{\plotone{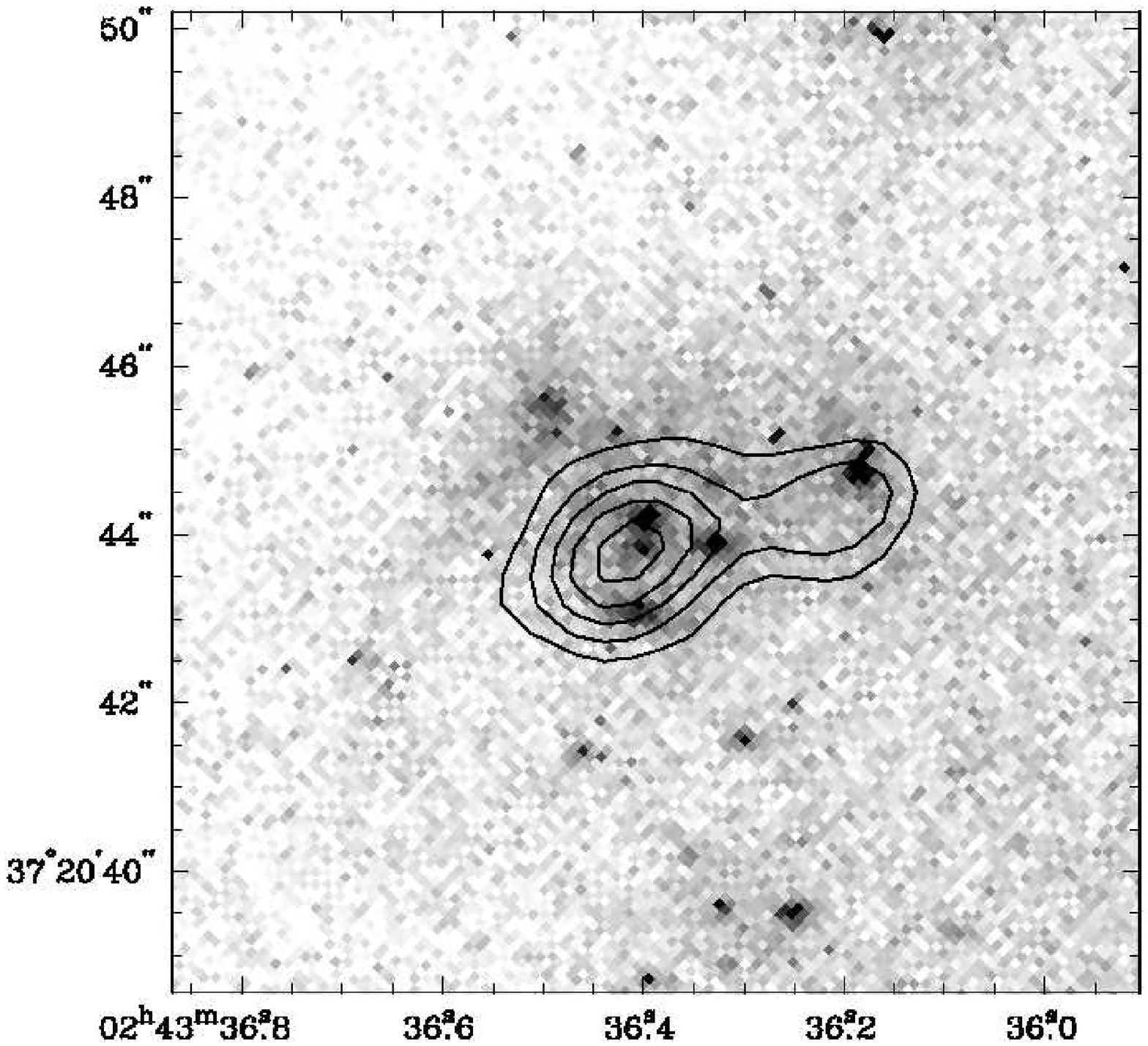}}
\caption{{\it HST} WFPC1 observations of SN\,1961V with the F675W filter 
(greyscale image) overlaid
with contours (0.07, 0.11, 0.15, 0.20, and 0.24 mJy) to show 6~cm radio
continuum emission from the VLA observations by \citet{stockdale01}. 
Coordinates given for the image are J2000.  The radio continuum 
emission shows the two nonthermal radio sources while the {\it HST}
shows the clusters in the centers of the two H{\sc ii} regions. Our 
echelle observations were centered on the H{\sc ii} region and thus 
may have missed the proposed optical counterpart for SN\,1961V.}
\label{fig_sn1961v}
\end{figure}

Our observations of SN\,1961V show two narrow-line ($\sim$30 \kms ) sources
in \ha\ and \oiii\ that are coincident with the two \ion{H}{2} regions,
but no broad H$\alpha$ emission line is detected in 2 hrs of integration 
time.  Furthermore, we do not detect \nii\ emission that might 
be expected if nitrogen enriched material in an LBV ejecta nebula 
is present.  It is possible that our slit just missed SN\,1961V.
As shown in Figure~\ref{fig_sn1961v}, the two radio sources are not 
exactly coincident with the two \ion{H}{2} regions.  The eastern radio
source, corresponding to SN\,1961V and coincident with the LBV
candidate identified by \citet{filip95}, is $\sim$2\arcsec\ southwest 
of the core of the eastern \ion{H}{2} region.  Since our E-W 
echelle slit was centered on the two \ion{H}{2} regions, whose
central star clusters were the only objects visible in the
telescope acquisition image, the 1\arcsec\ slit width could have easily 
missed SN\,1961V and the LBV candidate.

\subsection{SN\,1978K}

A host of lines are detected from SN\,1978K in NGC\,1313 (see 
Figure~\ref{fig_sn1978k}).  We clearly detect broad
\ha, \hb, and \ion{He}{1}\,$\lambda$5876 emission lines.  Furthermore,
we have a tentative weak detection of broad [\ion{O}{1}]\,$\lambda$6363 
emission.  Note that the stronger [\ion{O}{1}]\,$\lambda$6300 was not 
detected because
it has been Doppler shifted into a gap between echelle orders.  In addition
to broad emission lines, we detect narrow, spatially unresolved, emission
from \ion{He}{2}\,$\lambda$4686, \oiii\,$\lambda\lambda$ 4959,5007,
\nii\,$\lambda$5755, and \nii\,$\lambda\lambda$6548,6583.  
Interestingly, the \sii\,$\lambda\lambda$6716,6731 lines are not detected.
The implications of the line detections and non-detections from SN\,1978K
will be discussed further in \S~\ref{discuss}.

\subsection{SN\,1995N}
We detect broad \ha\ emission from SN\,1995N (see Figure~\ref{fig_sndetecta}),
with FWZI (full-width at zero intensity) of $\sim$3,000 \kms.  In addition,
a narrow interstellar component with FWHM $\sim$30 \kms\ is present.  While
the interstellar emission is red-shifted by roughly 40 \kms\ from the the 
systemic velocity of the host galaxy Arp\,261, the broad \ha\ emission 
appears to be blue-shifted by $\sim$200 \kms.

\subsection{SN\,1997eg}

Our observations of SN\,1997eg detect broad \ha\ emission with FWZI
of $\sim$3,500 \kms\ which is
blue-shifted by $\sim$400 \kms\ with respect to the host galaxy NGC\,5012.
Furthermore, these observations show narrow 
\ha\ and \nii\,$\lambda$6583 emission components and possibly
narrow \sii\,$\lambda\lambda$6716,6731 emission blue-shifted by 
$\sim$160 \kms\ with respect to the systemic velocity of the host galaxy.
The \ha\ and \nii\ emission are both narrow and are clearly
visible throughout the echelle slit and therefore are interstellar 
in origin.  
\citet*{sal02} observed SN 1997eg $\sim$1 yr earlier and detected 
a strong, narrow, P-Cygni profile caused by CSM,
which has completely vanished at the time of our observations.

\subsection{SN\,1998S}

The \ha\ spectrum of SN\,1998S in NGC\,3877 shows a total width 
greater than 10,000 \kms.  Superposed on this broad component from
the SN ejecta, we find a narrow \ha\ feature at $\sim$6581 \AA\ 
extending along the slit across the SN.  The spatial distribution
indicates the presence of ISM in the vicinity of SN\,1998S.
In addition we detect narrow, spatially extended, \nii\,$\lambda$6583
and narrow, spatially unresolved \oiii\,$\lambda\lambda$4959,5007 emission
(see Figure~\ref{fig_sn1998S}).
Comparison between the \ha\ and \nii\ lines on and off the SN position
does not show conclusively the existence of a circumstellar nebula.
The presence of the spatially unresolved \oiii\ emission 
indicates that there is a nebula associated with the SN progenitor.
The heliocentric velocities of the narrow \ha, \nii, and \oiii\ lines 
extracted at the position of the SN are $-$66, $-$74, and $-$64 \kms,
respectively, relative to the systemic velocity of the NGC\,3877.
The velocity difference between the \nii\ line and the other two lines
are larger than the uncertainty of the measurements, $\sim$3 \kms, indicating
that these narrow components may have a complex origin consisting of an
unknown mixture of CSM and ISM.

We can further compare our observations of SN\,1998S, taken $\sim$1 year after 
its discovery, to the high dispersion spectra taken 17.4 and 36.3
days after the discovery \citep{fassia01}.  Only the \oiii\ $\lambda$5007
line shows continuity from one epoch to the next.  The line shape changes
from an asymmetric profile on day 17.4, to a broader (FWHM$=$59$\pm$9 \kms) 
and symmetric profile on day 36.3, to a narrower (FWHM$=$37$\pm$3 \kms)
symmetric profile centered at the same velocity 1 year later.  At \ha,
the early nebular line had a P-Cygni profile, while our spectra show 
the line only in emission.  The most dramatic difference between the
early observations and our observations is in the \nii\
lines.  The \nii\ lines were not detected on day 17.4; the auroral \nii\
$\lambda$5755 line was observed to be at least 3 times as strong as the nebular
\nii\ $\lambda$6583 line on day 36.3; 1 year later, the nebular \nii\
line was detected but the auroral \nii\ line was not detected.  

The early line ratios indicate the presence of ionized gas denser than 
10$^6$ cm$^{-3}$ which can only be circumstellar in origin \citep{fassia01}.
If the narrow nebular features from all three epochs originate from the
same nebula, then our observations detect the outer, low--density region
of this circumstellar nebula.  On the other hand, we cannot rule out the
possibility that the narrow, spatially unresolved spectral features we 
observed have a significant contribution from the ISM.

\subsection{SN\,1998bm}
The galactic environment of SN\,1998bm is complex and it remains unclear 
whether it is in NGC\,2820 or NGC\,2820A.  SN\,1998bm itself is not detected, 
but the interstellar emission in the echellogram appears consistent with
that of an expanding shell of size $\sim$600 pc.  While this interstellar
emission might be from a supergiant shell, the
expansion velocity, $\sim$75 \kms, is relatively high compared to 
similar resolved structures in the Large Magellanic Cloud \citep{points99}
and other galaxies \citep{hunter97}.
We therefore conclude that the emission lines detected are likely
a superposition of the ISM of both NGC\,2820 and NGC\,2820A.
Each of these interstellar emission components is blue-shifted with respect
to the systemic velocity of the two galaxies.  The brighter component (toward
the western end of the slit) is probably associated with NGC\,2820A and is 
blue-shifted by $\sim$80 \kms.  It is less certain from which galaxy the fainter
component might arise but it is blue-shifted by either $\sim$120 \kms\ or 
$\sim$160 \kms\ from the systemic velocity of NGC\,2820 and NGC\,2820A, respectively.

\subsection{SN\,1999E}

Our observation of SN\,1999E detects faint,
spatially-unresolved, broad emission features throughout the
entire echellogram.  The high dispersion of the spectrum and
the gaps between echelle orders make it difficult to identify
these emission features.  It is nevertheless clear that these
emission features are unlike that of any stars; therefore, we believe 
that the broad emission features are from the young SN\,1999E.
In addition to the broad features, narrow interstellar
\ha, \hb, and \oiii\,$\lambda$5007 emission are detected.
The \ha\ emission detected has an average heliocentric velocity of 
7,925$\pm$10 \kms\
and FWHM of $\sim$100 \kms.  Both the \hb\ and \oiii\ lines have similar
redshifts and all lines have a velocity 185 \kms\ greater than the 
previously reported systemic velocity of the host galaxy 
IRAS\,13145$-$1817 \citep{ndc98}.

\subsection{SN\,1999Z}

Our observation of SN\,1999Z also detects faint,
spatially-unresolved, broad emission features throughout the entire
echellogram.  Similarly, we believe that these emission features 
originate from SN\,1999Z.  The H$\alpha$ line is red-shifted into a
gap between echelle orders but narrow interstellar components of 
H$\beta$, \oiii\,$\lambda$5007, and \sii\,$\lambda\lambda$6716,6731 
are detected.   Based on the two strongest lines, \hb\ and 
\sii\,$\lambda$6716, we find a heliocentric velocity of 
15,450$\pm$10 \kms\ for the emission lines which is $\sim$350 \kms\ 
greater than that previously reported for UGC\,05608 by \citet{jha1999}
which were based on the apparent velocity centroid of the then 
newly detected SN line emission.

\section{Discussion}\label{discuss}

The very last evolutionary stage of a massive star before 
SN explosion is not well known.
It is impossible to study SN progenitors after they have 
exploded, but it is possible to study the CSM
shed by the progenitors and photoionized by UV flashes from
the SNe.   The physical properties of these 
circumstellar nebulae may provide invaluable information 
about the doomed massive stars.

In the last hundred years, SN\,1987A has been the only
SN observed in the Local Group.  All the other SNe are in distant
galaxies, where circumstellar nebulae can only be detected 
in high-dispersion spectra of SNe as narrow nebular lines 
with high [\ion{N}{2}]/H$\alpha$ ratios.  The size of an 
unresolved circumstellar nebula can be determined if the SN is 
spectroscopically monitored until the SN ejecta impact the 
circumstellar nebula.  For example, if SN\,1987A were at a 
large distance, the size of its inner ring nebula could still 
be determined from the 10--11 years time lapse from the SN 
explosion to the emergence of a broader nebular component, 
and the 15,000 km~s$^{-1}$ expansion velocity of the SN ejecta 
inferred from its broad Ly$\alpha$ emission \citep{Mietal00}.

There are apparently two types of circumstellar nebulae 
produced by SN progenitors.   The first type of circumstellar 
nebulae are swept-up by SN ejecta within a year or two, as
demonstrated dramatically in time-sequenced spectra for 
SN\,1988Z \citep{stat91}.  In the case of SN\,1997eg, previous 
observations have shown a narrow P-Cygni component to the SN 
spectrum \citep{sal02} which has disappeared at the time of 
our observations, one year later.
These nebulae are small ($\sim$0.01 pc), and, for the case 
of SN\,1997eg, very dense ($\ge 10^7$ cm$^{-3}$).  
Such ionized nebulae have not been observed around any known 
evolved massive stars.  This circumstellar material may have been 
ejected immediately before the SN explosion \citep{Chu01} or 
during a red supergiant phase \citep{fassia01}.

The second type of circumstellar nebulae are longer lived, 
indicating a larger size.  These nebulae might be the 
counterparts of circumstellar nebulae observed around Wolf-Rayet 
(WR) stars or LBVs \citep{CWG99}.  For a nebular radius of 2 pc 
and a SN ejecta expansion velocity of 10,000 \kms, the 
impact of SN ejecta on the circumstellar nebula is expected at
$\sim$200 yr after the SN explosion, and the impact will produce 
X-ray-luminous SNRs such as the one observed in NGC\,6946 
\citep{dunne00} and the SNR 0540$-$69.3 in the Large Magellanic
Cloud \citep{CMB98}.

Among the SNe we surveyed, we detect CSM around SN\,1978K and
SN\,1998S.  In the case of SN\,1978K, the circumstellar nebula
was not yet hit by the SN ejecta 21 years after the SN explosion.
This must belong to the second type of circumstellar nebula
described above.  In the case of SN\,1998S, the rapid temporal
evolution of the spectral lines from the CSM over the first year
indicate that this material is interacting with the SN ejecta
and must belong to the first type.

\begin{figure}
\figurenum{6}
\centerline{\plotone{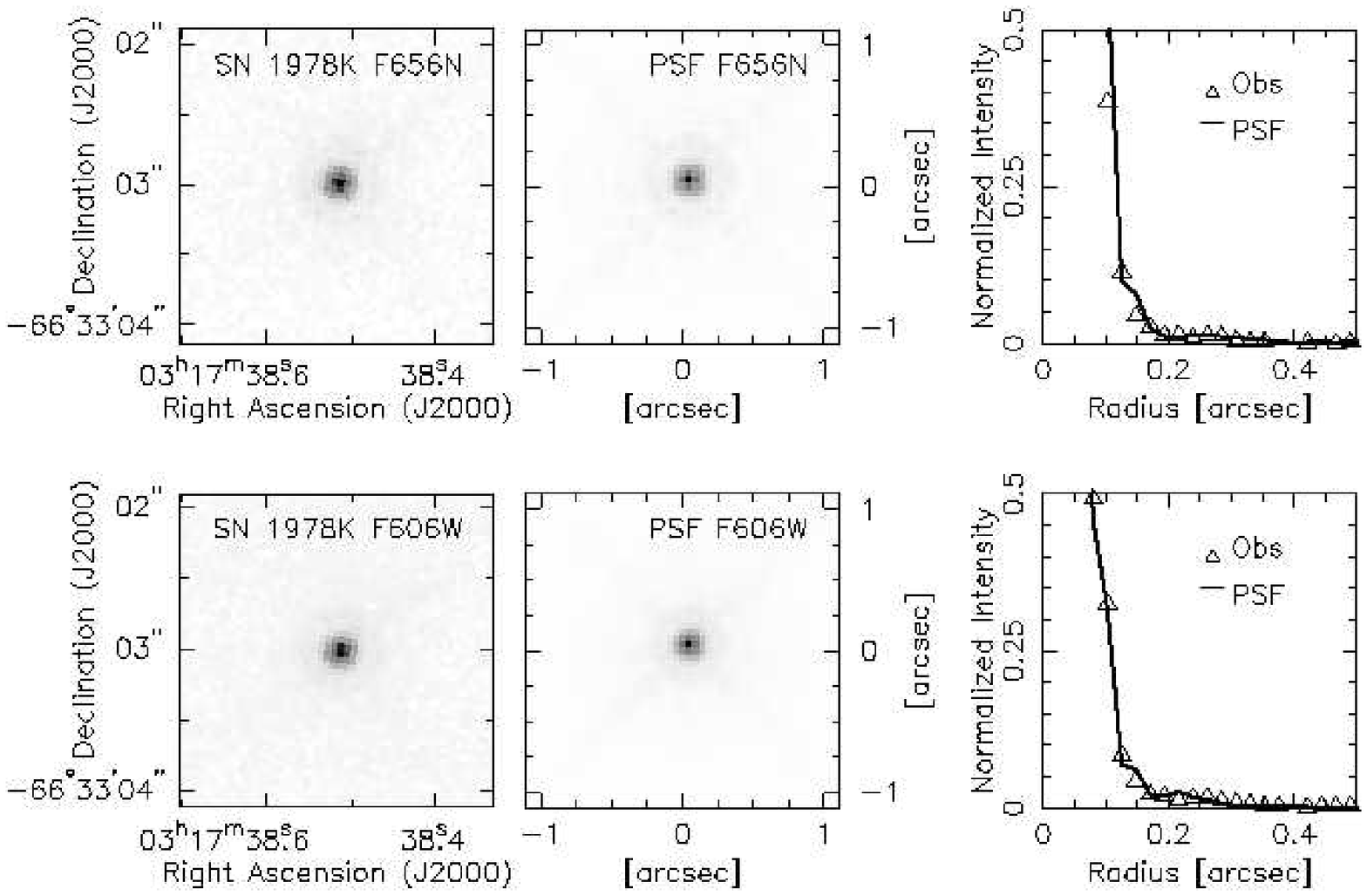}}
\caption{({\it Top left panel}) {\it HST} \ha\ (F656N) image of 
SN\,1978K, which was situated on the PC chip and is clearly unresolved. 
({\it Top center panel}) Synthetic PSF generated by TinyTim, assuming 
a monochromatic point source, for comparison.  ({\it Top right panel})
Average radial profile where the SN\,1978K F656N observations are 
plotted using triangles and the PSF generated by TinyTim is shown as 
a solid line.
({\it Bottom left panel}) Broad-band (F606W) image of SN\,1978K, also 
situated on the PC chip.  Note the source size is indistinguishable 
from the narrow-band image.  ({\it Bottom center panel})  Synthetic PSF 
generated by TinyTim for the F606W filter, using our echelle spectrum 
as input (i.e., no continuum emission). ({\it Bottom right panel)} 
Plot of average radial profile for the observations and synthetic PSF
for the F606W observations (triangles) and model (solid line).}
\label{fig_psf}
\end{figure}

Our new observations of SN\,1978K confirm the \ha\ and \nii\ line
emission and expansion velocity of the CSM previously reported 
by \citet{Chetal99}.  As this circumstellar nebula is in the
second class mentioned above and NGC\,1313 is relatively nearby,
the nebula might be large enough to be resolved by the {\it HST}.
Therefore, we obtained the {\it HST} archive WFPC2 images of 
SN\,1978K taken with F656N and F606W filters on 1998 September 23
as part of program GO-6713 (PI: W. Sparks).
In these observations the SN is centered in the PC and a point-like 
source is detected (see Figure~\ref{fig_psf}).  Also shown in 
Figure~\ref{fig_psf} are the expected point-spread-functions for 
position of the SN in the PC for observations with the F656N and 
F606W filters.\footnote{The theoretical PSFs were generated using 
Tiny Tim V5.0 written by John Krist and Richard Hook which is available 
from http://www.stsci.edu/software/tinytim.}
These images clearly place an upper limit of 0\farcs 1 on the
size of the circumstellar nebula of SN\,1978K.
At the distance of the host galaxy NGC\,1313, 4.5 Mpc \citep{devauc63}, 
this upper limit on the size corresponds to 2.2 pc.
This small size is more consistent with the sizes of LBV nebulae
than WR nebulae, as also concluded by \citet{Chetal99}.

In the case of SN\,1998S, our observations cannot unambiguously determine
whether the emission seen is circumstellar or interstellar.  In either
case, the nebular emission observed is no longer consistent with gas
denser than the critical density needed for the auroral \nii\ $\lambda$5575
line to dominate the nebular lines. Therefore, the density must be 
less than $\sim$10$^{5}$ cm$^{-3}$ and the emission seen is either
from a lower density outer region of a circumstellar nebula or from
interstellar material.  If the narrow lines we detected are indeed 
circumstellar in origin, then the blueward velocity offset of the \nii\
line relative to the \ha\ and \oiii\ lines may be caused by light travel
time effects similar to those used to explain the early evolution of the
\oiii\ line profile by \citet{fassia01}.  

Our high-dispersion (R$\sim$30,000) spectroscopic observations of 
SNe have demonstrated the possibility of detecting circumstellar
nebulae ejected by SN progenitors and deriving information on
the progenitors from the nebular properties.  
Our observations have also demonstrated the difficulty of such work 
because of the faintness of the SNe and their associated circumstellar
nebulae.  In order to identify emission from such nebulae, high-dispersion
spectroscopic observations of SNe with large telescopes such as Gemini and Keck 
are needed.  Only after a large number of such circumstellar 
nebulae have been detected and studied can we hope to better understand 
SN progenitors and their circumstellar environment.

\acknowledgements
We would like to thank the referee, E. Schlegel, for careful consideration
of our original manuscript and insightful comments.  In addition, we would
like to thank P. Lundqvist for useful discussion of SN\,1998S.

\clearpage

\begin{figure}
\figurenum{1b}
\centerline{\plotone{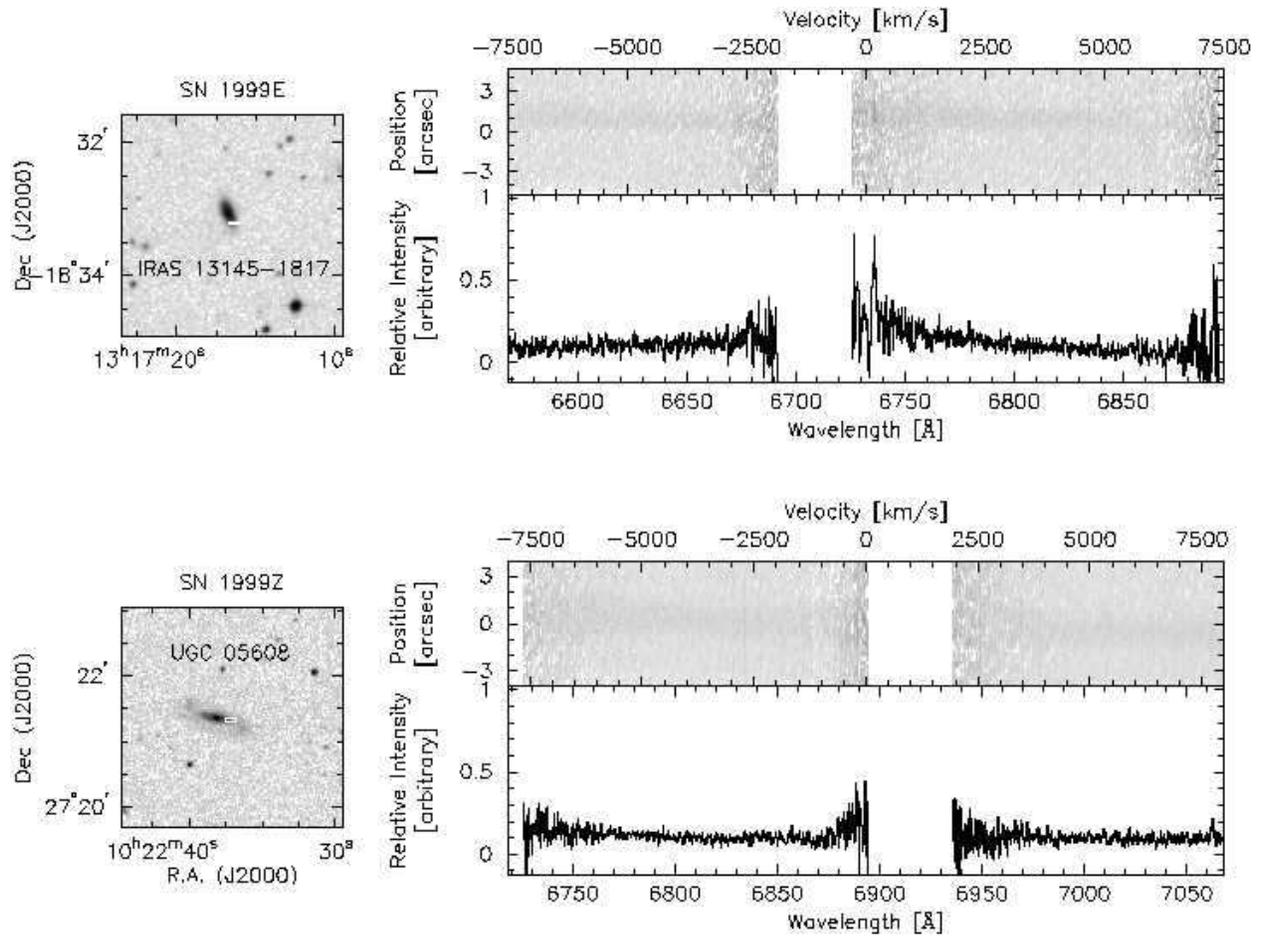}}
\caption{Same as Figure~\ref{fig_sndetecta} for SN\,1999E and SN\,1999Z.}
\label{fig_sndetectb}
\end{figure}

\clearpage 

\begin{deluxetable}{lcr}
\tablewidth{0pt}
\tablecaption{Journal of Observations\label{tab_obssum}}
\tablehead{
\colhead{Object} & \colhead{Date} & \colhead{Exposures\tablenotemark{a}}
}
\startdata
SN\,1954J    & 1999 Mar 3 & 1$\times$1800 \\
SN\,1961V    & 1999 Mar 2 & 1$\times$3600 \\
             & 1999 Mar 3 & 2$\times$1800 \\
SN\,1978K\tablenotemark{b} 
             & 2000 Dec 5 & 3$\times$\phn900  \\
SN\,1979C    & 1999 Mar 3 & 1$\times$1800 \\
SN\,1992ad   & 1999 Mar 3 & 1$\times$1800 \\
SN\,1993J    & 1999 Mar 2 & 1$\times$1800 \\
SN\,1995N    & 1999 Mar 3 & 2$\times$1800 \\
SN\,1996bu   & 1999 Mar 2 & 1$\times$1800 \\
SN\,1997ab   & 1999 Mar 3 & 1$\times$1800 \\
SN\,1997bs   & 1999 Mar 4 & 1$\times$1800 \\
SN\,1997dn   & 1999 Mar 3 & 1$\times$1800 \\
SN\,1997eg   & 1999 Mar 2 & 2$\times$1800 \\
SN\,1998S    & 1999 Mar 2 & 1$\times$1800 \\
             & 1999 Mar 3 & 3$\times$1800 \\
SN\,1998bm   & 1999 Mar 2 & 2$\times$1800 \\
SN\,1999E    & 1999 Mar 3 & 1$\times$\phn900  \\
SN\,1999Z    & 1999 Mar 4 & 2$\times$1800 \\
\enddata 

\tablenotetext{a}{Exposures are denoted as (number of exposures)$\times$
(exposure time in seconds for each exposure).}
\tablenotetext{b}{Observations obtained with the CTIO 4m telescope 
(all other observations were obtained with the KPNO 4m telescope).}

\end{deluxetable}

\begin{deluxetable}{lllcccc}
\tablewidth{0pt}
\scriptsize
\tablecaption{Summary of Observations\label{tab_observe}}
\tablehead{
\colhead{}       & \colhead{} & \colhead{}   & 
\colhead{Systemic} & \multicolumn{3}{c}{}   \\
\colhead{}       & \colhead{SN} & \colhead{Host}   & 
\colhead{Velocity\tablenotemark{a}} & \multicolumn{3}{c}{H$\alpha$ Detected from} \\
\cline{5-7}
\colhead{Object} & \colhead{Type} & \colhead{Galaxy} & 
\colhead{[km s$^{-1}$]} & 
\colhead{SN} & \colhead{ISM} & \colhead{CSM} 
}
\startdata
SN\,1954J   & {\sc II}pec & NGC\,2403          & \phs\phm{/}\phn\phn\phn131  &  no & yes & no\\
SN\,1961V   & {\sc II}pec & NGC\,1058          & \phs\phm{/}\phn\phn\phn518  &  no & yes & no\\
SN\,1978K   & {\sc II}pec & NGC\,1313          & \phs\phm{/}\phn\phn\phn475  & yes &  no & yes\\
SN\,1979C   & {\sc II}-L  & NGC\,4321          &    \phs\phm{/}\phn\phn1571  &  no & yes & no\\
SN\,1992ad  & {\sc II}    & NGC\,4411b         &    \phs\phm{/}\phn\phn1270  &  no & yes & no\\
SN\,1993J   & {\sc II}b   & NGC\,3031          &\phm{/}\phn\phn\phn\phn$-$34 &  no &  no & no\\
SN\,1995N   & {\sc II}    & Arp\,261           &    \phs\phm{/}\phn\phn1834  & yes & yes & no\\
SN\,1996bu  & {\sc II}pec & NGC\,3631          &    \phs\phm{/}\phn\phn1156  &  no &  no & no\\
SN\,1997ab  & {\sc II}n   & HS\,0948$+$2018    &    \phs\phm{/}\phn\phn3598  &  no &  no & no\\
SN\,1997bs  & {\sc II}pec & NGC\,3627          & \phs\phm{/}\phn\phn\phn727  &  no & yes & no\\
SN\,1997dn  & {\sc II}    & NGC\,3451          &    \phs\phm{/}\phn\phn1334  &  no & yes & no\\
SN\,1997eg  & {\sc II}pec & NGC\,5012          &    \phs\phm{/}\phn\phn2619  & yes & yes & no\\
SN\,1998S   & {\sc II}    & NGC\,3877          & \phs\phm{/}\phn\phn\phn902  & yes & yes & yes\\
SN\,1998bm  & {\sc II}    & NGC\,2820/20A      &                \phs1534/74  &  no & yes & no\\
SN\,1999E   & {\sc II}n   & IRAS\,13145$-$1817 & \phs\phm{/}\phn\phn7925\tablenotemark{b}  & likely & yes & no\\
SN\,1999Z   & {\sc II}n   & UGC\,05608         &    \phs\phm{/}\phn15453\tablenotemark{b}  & likely & yes\tablenotemark{c} & no\\
\enddata 
\tablenotetext{a}{Systemic velocities of the host galaxies are given in 
the heliocentric frame and come from \citet{tully88}.}
\tablenotetext{b}{The observed heliocentric velocity of the ISM component
from this paper.  These values are likely more accurate for the systemic
velocity of the galaxy than previous values in the literature.}
\tablenotetext{c}{The detection of emission from the ISM of UGC\,05608 is 
based on \hb, \sii, and \oiii\ emission but not \ha.}
\end{deluxetable}

\end{document}